\documentstyle[12pt, a4]{article}

\begin{document}

\author{B. G\"{o}n\"{u}l\footnote{gonul@gantep.edu.tr},
O. \"{O}zer and M. Yilmaz \and Department of Engineering
Physics, University of Gaziantep, \and 27310 Gaziantep-T\"{u}rkiye}
\title{Investigation of phase-equivalent potentials by a halo transfer reaction}
\date{}
\maketitle

\begin{abstract}
Using the supersymmetric quantum mechanics we investigate the wave
function-sensitive properties of the supersymmetric potentials which have
received a lot of attention in the literature recently. We show that a
superdeep potential and its phase-equivalent shallow-partner potential give
very similar $rms$ values for the weakly bound systems such as the deuteron
and $^{11}Be$ nuclei. Although the corresponding eigenstates differ in the
node-number, our investigation on the $^{11}Be(p,d)^{10}Be$ single nucleon
halo transfer reaction at 35 MeV show that also other physical quantities
such as the cross section angular distributions calculated using these wave
functions reflect the nodal structure rather weakly. This lends support to
two nearly equivalent treatments of the Pauli principle.
\end{abstract}

{\it Published in} Eur. Phys. J. A {\bf 9}, 19-28 (2000)

\section{Introduction}

The long-standing dichotomy of choosing a shallow or deep effective local
potential to describe nucleus-nucleus elastic scattering was greatly
clarified by Baye \cite{Baye1}, who demonstrated that in scattering, for all
practical purposes, these two kinds of potentials in the literature \cite
{Ali&Bodmer1},\cite{Buck1} are phase-equivalent supersymmetric partner
potentials of each other. Supersymmetric quantum mechanics allows one to
transform a hamiltonian to its partner such that they possess identical
spectra or differ at most from each other by having the lowest eigenstate
eliminated in its partner's spectrum \cite{Adrianov}. Repeated application
of the supersymmetric transform could yield a hamiltonian which has a
prescribed number of eigenstates less than the starting hamiltonian. By the
same token, a supersymmetric transform can also add an eigenstate of the
desired energy to the starting spectrum. The supersymmetric transform, apart
from eliminating the lowest eigenstate, induces in the first instance a
change in the phase shifts of the continuum states of the starting spectrum.
Baye \cite{Baye1} showed that by a judicious choice of repeated
supersymmetric transforms, the partner hamiltonian can be made to be
phase-equivalent. Thus, in the case of $\alpha -\alpha $ scattering, where
the starting hamiltonian has a deep effective local potential \cite{Buck1}
known to possess bound states which are unphysical because of the Pauli
principle, Baye eliminated these unphysical states by successive
supersymmetric transform while preserving the phase equivalence. In the
course of this process, the original potential is transformed to be singular
and progressively shallower. Moreover, the supersymmetric transforms are
performed for each angular momentum ($\ell $) separately. Therefore, at the
final stage, a set of shallow, energy-independent and $\ell $-dependent
effective local potentials is obtained which is completely phase-equivalent
to the original deep, energy-independent and $\ell $-independent potential.
The set of shallow potentials bears a remarkable resemblance to the $\ell $%
-dependent $\alpha -\alpha $ potentials documented in the literature \cite
{Ali&Bodmer1}.

Such ambiguous choice in the nature of the potential to describe
nucleus-nucleus elastic scattering is found not only in $\alpha -\alpha $
but also in many other systems \cite{Baye2}. This dual picture of the
nucleus-nucleus interaction arises most probably because the many-body
description of the system has been simplified in different ways to a
two-body interaction between two structureless particles. For the analysis
of elastic scattering, such drastic difference in the character of the
potentials is immaterial since only phase shifts are required and these
potentials are phase-shift equivalent or in the main so. However, when one
has to choose one of these potentials to be used in nuclear structure
studies in which wave functions are explicitly involved, the implication of
a deep or shallow potential may be immense. The supersymmetric procedure
produces two sets of wave functions for weakly bound systems, which coincide
at large distances but differ at small distances by the additional node
appearing inside the core by use of the deep potential. Therefore, it is
important to have a quantitative criteria to discriminate which potential
has the correct wave function-sensitive properties.

These deep and shallow potentials were used independently by Baye {\it et al.%
} and Liu \cite{Liu} to calculate bremsstrahlung emission in possible
nucleus-nucleus collisions. The calculations seemed to indicate that while
the bremsstrahlung cross sections from resonating-group method and the deep
potential resemble each other, those of the shallow potential are distinctly
different. Hence it was concluded that the deep potential is to be preferred
over its shallow partner if wave function-sensitive properties are important.

Recently Dijk {\it et al.} \cite{Van Dijk}, and Ridikas and his co-workers
\cite{Van Dijk}, have separately shown that a superdeep potential and its
supersymmetric partner give very similar $rms$ values for the model
deuteron, and one-neutron halo systems (considering $^{11}$Be nucleus),
respectively. In the work of Ridikas {\it et al.} it was also stated that
other physical quantities, which are more sensitive to the behavior of the
radial wavefunctions in the nuclear interior, such as transition
probabilities obtained by the deep potential and its phase equivalent
partner, reflect the nodal structure rather weakly.

In order to complete our discussion, we repeat these calculations
with some extent, considering a transfer reaction,
$^{11}Be(p,d)^{10}Be$, involving the weakly bound deuteron and
single nucleon-halo $^{11}Be$ nuclei which are well suited for
studying the consequence of different wave functions from the deep
and reconstructed phase equivalent shallow potentials on reaction
observables. Current experimental activity in the area of
light-neutron rich and drip-line nuclei now dictates the rapid
development of calculable theoretical models for reactions and
scattering of effective few-body systems. Hence, there is an
increasing general interest in supersymmetric potentials in this
context. Our results are contributions to the discussion in this
subject -an investigation in a relatively unexplored area of the
quantitative consequence of the supersymmetric potentials.

We begin with a brief sketch of the general method in section 2, where we
also introduce a two-parameter, deep, shape invariant two-body potential
used throughout the present calculations. In sections 3 and 4, we discuss
the application of the method to the deuteron and $^{11}Be$ nuclei
respectively, giving the characteristic properties of the constructed
phase-equivalent two-body potentials, and the connection between the
exclusion of deep-lying Pauli forbidden bound states from some potential and
supersymmetry is reviewed in the light of the calculation results. Section 5
discusses the $^{11}Be(p,d)^{10}Be$ reaction calculations in terms of the
supersymmetric partner potentials. Finally section 6 contains a summary and
the conclusion.

\section{Supersymmetric quantum mechanics}

Supersymmetric quantum mechanics \cite{Witten} and its connection to the
factorization method \cite{Infeld} have been extensively investigated \cite
{Cooper}. Since the ground state wave function $\Psi ^{(n=0)}$ for a bound
system has no nodes, it can be written as

\begin{equation}
\Psi ^{(n=0)}(r)=\exp (-\frac{\sqrt{2\mu }}\hbar \int W(r)dr)
\end{equation}

Introducing the operators
\begin{equation}
\widehat{B}=W(r)+\frac
{i}{\sqrt{2\mu}}\hat{p}~,~\widehat{B}^{+}=W(r)-\frac
{i}{\sqrt{2\mu }}\hat{p}
\end{equation}
the hamiltonian can be easily factorized

\begin{equation}
\hat{H}-E^{(n=0)}={\widehat{B}^{+}}\widehat{B}
\end{equation}
where $E^{(n=0)}$ is the ground state energy. Since the ground state wave
function satisfies the condition
\begin{equation}
\widehat{B}\left| \Psi ^{(n=0)}\right\rangle =0
\end{equation}
the supersymmetric partner hamiltonians ($H_{m}$, $m=1,2,...$)
\begin{equation}
{\hat{H}}_1={\widehat{B}^{+}}{\widehat{B}}~~,
~~\hat{H}_{2}=\widehat{B}\widehat{B}^{+}
\end{equation}
have the same energy spectra except the ground state of
$\hat{H}_{1}$, which has no corresponding state in the spectra of
$H$. The corresponding supersymmetric partner potentials are given
by
\begin{equation}
V_{1}(r)=\left[W(r)\right]^{2}-\frac{\hbar}{\sqrt{2\mu}}\frac{dW}{dr}
~,~
V_{2}(r)=\left[W(r)\right]^{2}+\frac{\hbar}{\sqrt{2\mu}}\frac{dW}{dr}
\end{equation}

It was shown that a subset of the potentials for which the Schr\"{o}dinger
equations are exactly solvable share an integrability condition called shape
invariance \cite{Gendenshtein}. The partner potentials of Eq. (6) are called
shape invariant if they satisfy the condition
\begin{equation}
V_{2}(r;a_{1})=V_{1}(r;a_{2})+R(a_{1})
\end{equation}
where $a_{1,2}$ are a set of parameters that specify
space-independent properties of the potentials (such as strength,
range, and diffuseness), $a_{2}$ is a function of $a_{1}$, and the
remainder $R(a_{1})$ is independent of $r$.

An iterative procedure within the supersymmetric quantum mechanics
framework for building the partner of a given potential admitting
the same eigenvalues except for that of the missing ground state
was proposed by Baye \cite{Baye1} on the basis of a general
procedure due to Sukumar \cite{Adrianov}, \cite {Sukumar}. The
method relies on a factorization property of the hamiltonian, and
makes possible the (exact) construction of the partner potential
starting from the original potential ground state wave function.
It actually requires two steps, the intermediate potential
$(V_{2})$ having the same negative energy spectrum for a bound
system as the original potential $(V_{1})$, except for the ground
state of the latter, but a different phase shift; $V_{2}(r)$ is
given by
\begin{equation}
V_{2}(r)=V_{1}(r)-2\frac{\hbar^{2}}{2\mu }\frac{d^{2}}{dr^{2}}\ln
\Psi_{1}
\end{equation}
where $\Psi_{1}(E_{1}^{(n=0)})$ denotes the original ground state
wave function. The second step provides $V_{3}(r)$ the final
phase-equivalent potential (PEP) as
\begin{equation}
V_{3}(r)=V_{1}(r)-2\frac{\hbar^{2}}{2\mu }\frac{d^{2}}{dr^{2}}\ln
\left[\Psi_{1}(E_{1}^{(n=0)})\Psi_{2}(E_{1}^{(n=0)})\right]
\end{equation}
where $\Psi_{2}(E_{1}^{(n=0)})$ stands for the wave function at
the same energy, but calculated with the intermediate potential
$V_{2}$. Eq. (9) can also be reduced to the form
\begin{equation}
V_{3}(r)=V_{1}(r)-2\frac{\hbar^{2}}{2\mu
}\frac{d^{2}}{dr^{2}}\left\{ \ln \int_0^r\left[ \Psi
_{1}(E_{1}^{(n=0)},r^{\prime })\right]^{2}dr^{\prime }\right\}
\end{equation}

Elimination of more than one state is accomplished by iterating this
two-step procedure.

\subsection{A two-parameter superdeep potential}

In Ref.\cite{Michel}, the on-shell equivalence of the deep
quantum-chromodynamically motivated realistic nucleon-nucleon interaction
proposed by Kukulin {\it et al.} \cite{Kukilin} with more conventional
repulsive-core forces has been investigated by eliminating its unphysical
deeply bound states, while preserving its scattering properties and the
binding energy of the deuteron. Using the spirit of this work, and of Ref.%
\cite{Van Dijk}, here we use an alternative superdeep potential. As a
simple, physically interesting example, consider the potential
\begin{equation}
V(r)=-V_{0}sech^{2}\beta r
\end{equation}

Potentials of this shape can be generated from the superpotential
\cite{Dutt},
\begin{equation}
W(r)=A\tanh \beta r,~~A>0.
\end{equation}
In fact, using Eq. (6), the supersymmetric partner potentials are
\begin{eqnarray}
V_{1}(r;A)=A^{2}-A\left( A+\frac{\beta \hbar }{\sqrt{2\mu
}}\right)sech^{2}\beta r~,  \nonumber \\ V_{2}(r;A)=A^{2}-A\left(
A-\frac{\beta \hbar }{\sqrt{2\mu }}\right) sech^{2}\beta r~.
\end{eqnarray}
Clearly, one can write
\begin{equation}
V_{2}(r;A)=V_{1}(r;A-\frac{\beta \hbar }{\sqrt{2\mu
}})+A^{2}-A\left( A-\frac{\beta \hbar }{\sqrt{2\mu }}\right)^{2}~,
\end{equation}
which is precisely the requirement of Eq. (7) for shape
invariance. Therefore the bound state energies of the potential
$V_{1}$ are
\begin{equation}
E_{1}^{(n)}=A^{2}-\left( A-n\frac{\beta \hbar }{\sqrt{2\mu
}}\right)^{2}~,
\end{equation}

The energy levels $E^{(n)}$ of the original potential given by Eq.
(11) can be obtained by subtracting $A^{2}$ from $E_{1}^{(n)}$ and
identifying
\begin{equation}
V_{0}=A\left( A+\frac{\beta \hbar }{\sqrt{2\mu }}\right)
\end{equation}
Solving for $A$ and requiring $A>0$ gives
\begin{equation}
A=-\frac{\beta \hbar }{2\sqrt{2\mu
}}+\frac{1}{2}\sqrt{\frac{\beta^{2}\hbar^{2}}{\sqrt{2\mu
}}+4V_{0}}~.
\end{equation}
Therefore, the energy levels of the deep potential
$V(r)=-V_{0}sech^{2}\beta r$ are
\begin{equation}
E^{(n)}=E_{1}^{(n)}-A^{2}=-\left( A-\frac{n\beta \hbar
}{\sqrt{2\mu }}\right)^{2}~~,
\end{equation}
which is well known to be the correct answer \cite{Landau}. As we deal with
the bound systems, we require the odd solutions due to boundary conditions.
Hence replacing $n$ in Eq. (18) by $2n+1$ term, we arrive at ,
\begin{equation}
E^{(n)}=-\frac{\hbar ^{2}}{2\mu }({\tilde{A}}-2n-1)^{2}\beta
^{2}~~,~~n=0,1,2,...~,
\end{equation}
where ${\tilde{A}}=\frac {A}{( \hbar \beta /\sqrt{2\mu })}$. The
depth of the potential given by Eq. (16) reduces in this case to
the form
\begin{equation}
{\tilde{V}}_{0}=-\frac{\hbar ^{2}}{2\mu
}{\tilde{A}}({\tilde{A}}+1)\beta ^{2}
\end{equation}

Eqs. (19) and (20) are in consistent with the expressions used in
Ref. \cite {Van Dijk} where also the analytical expressions for
the wave functions of the ground and first excited state
corresponding the potential of interest can be found.

We have first employed this deep sech-squared potential in
analyzing the alpha-alpha scattering (but not discussed here) by
choosing the two parameters as ${\tilde{A}}=5.945$ and $\beta
=0.535$ $fm^{-1}$, together with $\frac{\hbar ^{2}}{2\mu _{\alpha
-\alpha }}=10.375$ MeV $fm^{2}$. We have observed that
$-V_{0}sech^{2}\beta r\rightarrow -U_{0}exp^{(-\alpha r^2)}$ with
$U_{0}=122.694$ $MeV$, $\alpha =0.22$ $fm^{-2}$ and have
reproduced successfully the Figs. (1,2) of Ref. \cite{Baye1} using
this two-parameter shape invariant superdeep potential, without
involving a gaussian type potential in the calculations. However,
as we deal with $^{11}Be(p,d)^{10}Be$ reaction calculations
throughout the present work, we focus on the treatment of deuteron
and $^{11}Be$ ground state wave functions in the supersymmetric
quantum mechanical framework.

\section{Application to the deuteron system}

All the available ''realistic'' nucleon-nucleon forces are
characterized by a relatively weak central attractive part and by
the presence of a hard or soft repulsive core at small distances;
the first feature reflects the loose binding of the neutron-proton
system, while the introduction of a repulsive core is required by
the negative values assumed by the low-experimental phase-shifts
when energy increases. However the feasibility of a description of
comparable quality of the low energy properties of the two-nucleon
system in the $^{1}S_{0}$ and $^{3}S_{1}$ - $^{3}D_{1}$ channels
(including deuteron properties), in terms of a deep, purely
attractive interaction called Moscow potential was demonstrated by
Kukulin and his co-workers \cite{Kukilin}. Their potential differs
from those obtained in the more traditional approaches by the
existence of an additional deeply-bound state in each channel, and
by an increase of the absolute singlet and triplet phase-shifts
due to this extra unphysical bound state. It is well known from
cluster nuclear physics that these seemingly contradictory
features - that is, repulsive core versus deep potential
descriptions - are two ways to simulate the effects of the Pauli
principle in a local potential model description when the two
interacting particles are composed of identical fermions. There
have been quite a few attempts to derive the features of the
two-nucleon interaction from a quark picture of the nucleon. The
most non-relativistic quark model calculations led to an effective
nucleon-nucleon interaction with a strong repulsive core and an
intermediate range attraction similar to those displayed by the
empirical potentials. On the other hand, the work described in
Ref.\cite{Kukilin} indicated, for the nucleon-nucleon scattering,
that the relative s-wave function has to have a node at small
distance. The existence of a node in the relative motion wave
function can readily incorporated in a local potential
description, provided interaction is deep enough to accommodate
one (nodeless) deeply bound state, such as the one proposed by
Kukulin and his co-workers. The work described here will, in
addition to the other investigations undertaken, demonstrate
explicitly the equivalence of such deep potentials with the more
orthodox repulsive core empirical interactions, by constructing
the phase-equivalent supersymmetric partner of the deep potential
freed from unphysical bound states but still binding the deuteron
with correct energy. The resulting supersymmetric potential for
the deuteron case are to be shown to have a short range repulsive
core followed by a shallow attractive part, which are very similar
to those displayed by realistic interactions such as the Reid soft
core potential \cite{Reid}.

\subsection{Phase-equivalent potentials for the deuteron}

Now, superdeep potentials such as the Moscow potential give deuteron wave
functions with a node at short distances. The node arises because there is
an additional bound state which is Pauli forbidden for the actual
neutron-proton ($n-p$) system. The latest version of Moscow potential \cite
{Kukilin} includes both central and a tensor component, together with the
central and tensor one-pion exchange contributions (OPEP). In addition, the
equations in Ref.\cite{Kukilin} to solve both for the bound state ($S$- and $%
D$-wave) and the scattering problem in case of triplet potential are more
complicated, one needs to work out the usual coupled equations. For
simplicity, in present calculations we use an alternative superdeep
potential discussed in Section 2.1, which produces a model deuteron wave
function that has a node like the Moscow potential. However, this simple
potential does not have the required OPEP tail and do not consider $D$-wave
related to the tensor potential. So the physical observables calculated by
this potential, such as the radius of the deuteron which we will deal with
later in this section, should not be compared with the experimental value.
We note at this point that our aim here is not the rigorous reproduction of
experimental data but to test the reliability of a deep potential
description involving some unphysical bound states, and in particular to
search for the wave function sensitivity features of the deep and of its
supersymmetric partner-shallow potentials used both in the entrance and exit
channels of the $^{11}Be(p,d)^{10}Be$ reaction. For this reason, the use of
an appropriate simple potential, such as the binary $\sec $h-squared
potential, in the present analysis does not cause any physical problem.

Considering the well-known charge radius formula
\begin{equation}
R^{2}(charge)=\frac {1}{2}R_{p}^{2}+\frac {1}{4}R_{rms}^{2}
\end{equation}
with $R_{p}$ being the proton radius and $R_{rms}$ the mean-square
intercluster distance (matter radius), one can determine the free
parameters ${\tilde{A}}$ and $\beta $ for the potential considered
in analyzing the deuteron nucleus by solving the following system
equations ;
\begin{eqnarray}
E^{(n)}&=&-\frac{\hbar ^{2}}{2\mu
_{n-p}}({\tilde{A}}-2n-1)=-2.226~MeV~, \\ R_{rms}^{2} &=&\frac
{1}{4}\int drr^{2}\Psi _{m}^{(n)}({\tilde{A}},\beta
,r)^{2}=(1.95~~fm)~, \nonumber
\end{eqnarray}
where $n$ denotes the energy level as stated earlier and $m$ refers to $%
m^{th}$ hamiltonian. Throughout our calculations, $\frac{\hbar
^{2}}{2\mu_{n-p}}$ is set $41.47$ $MeV$ $fm^{2}$ and the arbitrary
constants ${\tilde{A}}$ and $\beta $, for the Moscow-type binary
potential, are calculated as $3.146$ and $1.587$ $fm^{-1}$
respectively. In this case, from Eq. (22), the ground state has a
binding energy of about $481$ $MeV$, which is unphysical and needs
to be suppressed. The physically meaningful deuteron bound state
for this superdeep potential corresponds to the first excited
state having a binding energy of $2.226$ $MeV$.

The building the partner of a given potential admitting the same
eigenvalues except for that of the missing ground state has been
discussed in Section 2. The two-parameter superdeep
potential-$V_{1}(r)$ and its phase-equivalent supersymmetric
partner-$V_{3}(r)$, together with the intermediate
non-phase-equivalent potential $V_{2}(r)$, are shown in Fig. 1-a,
and their corresponding wave functions in Fig. 2. As a result of
the presence of one spurious bound state, the deuteron wave
function for the superdeep potential possesses one node near the
origin (around $0.56$ $fm$). Figure 1b compares the phase
equivalent repulsive core interaction ($V_{3}$) with the central
part of the usual realistic Reid soft core potential \cite{Reid}.
The general futures of the PEP (such as the radius of the
repulsive core and the strength of the attractive part) are seen
to be similar to those of the Reid soft core potential. In spite
of different analytical behavior near the origin for both Reid
Soft core interaction (behaves at $r\rightarrow 0$ as $
e^{-(const\times x)}/x$) and the transformed phase equivalent
shallow potential ($V_{3}(r)_{r\rightarrow 0}\approx
(const/r^{2})$), we observe in the figure a considerable
similarity of both interactions. This means we have very tight
interrelation between a deep nucleon-nucleon model potential and
the standard Reid soft core interaction, which will be discussed
later in this section.

It is seen from fig. 2 that our reconstructed phase-equivalent
supersymmetric partner potential ($V_{3}$) has led to relative
motion wave functions very similar to this generated by the deep
potential ( $V_{1}$) outside the core region, but which lack the
small distance radial node resulting from the suppression of the
unphysical bound state. If there is a node in the wave function
and the wave function is reasonably large at small distances, then
one might expect that because of the normalization the wave
function at large distances would be reduced. In other words, the
asymptotic phase-equivalent wave function will have a smaller
value of the asymptotic amplitude and hence the radius would be
reduced. To clarify that if these wave functions having different
behavior inside the core lead to quantitatively different results,
we investigate the dependence of radius calculations, as an
observable, on the wave function properties.

The deuteron matter radius $R_{rms}$ can be calculated numerically
from Eq. (22) for either $\Psi (r)=\Psi _{1}^{(n=1)}(r)$ or $\Psi
(r)=\Psi _{3}^{(n=0)}(r) $, the bound state wave functions for the
superdeep and
transformed partner potentials. The numerical calculations of $rms${\it \ }%
value (root-mean-square radius) show that the radius of deuteron is $1.953$ $%
fm$ and $1.955$ $fm$ for the superdeep potential and PEP respectively, which
are so close. That means as a physical quantity, the radius calculated using
the wave functions having one-/no-node reflect the nodal structure rather
weakly. But, in case of experiment requiring a reduction in $R_{rms}$ for a
nucleus then short range contribution to the potential appears, such as the
superdeep potential used here, would be necessary.

In order to check the accuracy of the supersymmetric quantum mechanical
methods used in constructing the phase-equivalent potential, we have carried
out additional calculations on the phase-shift. The resulting phase-shift
curve obtained by PEP ($V_{3}(r)$) is compared with that obtained by the deep (%
$V_{1}$) potential and an excellent agreement between the
resulting and initial phase-shifts is observed.

To summarize, when nucleons are endowed with quark structures,
nucleon-nucleon
bound and scattering properties can be described by a deep potential ($%
\approx 1000$ $MeV$), whose supersymmetric partner potential is singular and
looks teasingly like the Reid soft-core or similar shallow potentials. One
can make a conclusion at this stage on the supersymmetry aspect of
nucleon-nucleon interaction in the light of the calculation results
obtained, making a possible connection between the exclusion of deep-lying
bound states from some potential and supersymmetry. This very interesting
aspect is connected with very deep interrelation between many-body and
potential treatment for the composite particle interaction in case of the
particles composed of elementary fermions. From physical point of view this
interrelation means the existence of a deeply hidden relation between the
relative motion of composite nucleus comprised from fermions and the
internal excitations of the composites. In fact, in non-relativistic quantum
mechanics which is used for the treatment of interaction of the composite
particles, the relative motion of the composites is treated as a bosonic
degree of freedom (i.e., no Pauli principle constrains are put to the
relative motion). On the other hand, the internal excitations of the quark
degrees of freedom inside the composites in the process of the mutual
collision of the composites should be treated as the manifestation of
fermionic degrees of freedom. The main problem in description of composite
particle interaction is the complicated interrelation between relative
motion of the composites and their inner excitations. And from this point of
view, the existence of the above supersymmetry aspect could mean that the
collision of such composites should be described correctly only within the
framework of supersymmetrical quantum mechanics and is only the projection
of this nontrivial picture onto mutual relative motion of the composites.

As a conclusion, by means of supersymmetric quantum mechanics we have found
that the two-parameter superdeep potential and its supersymmetric partner,
which is phase-equivalent to the former and freed from the unphysical deeply
bound states, give very similar $rms$ values for the deuteron system.
Although the corresponding eigenstates differ in the node-number, our
investigations have shown that the matter radius calculations using these
wave functions reflect the nodal structure rather weakly. This lends support
to two nearly equivalent treatments of the Pauli principle by choosing the
physical solution either by node-number criteria or by inclusion of a
repulsive part of the potential at the origin. However, the similar $rms$
results obtained do not automatically imply that other observables such us
differential cross section, vector analyzing power calculated using these
wave functions have to coincide. Clearly considerable additional work is
still needed to test further the virtues of a deep potential description of
the nucleon-nucleon interaction, which will be discussed in detail in
Section 5 considering a halo transfer reaction. However, as the $^{11}Be$
nucleus in the entrance channel of the reaction considered is weakly bound,
like the deuteron in the exit channel, a similar discussion for the $^{11}Be$
system being one-neutron halo nucleus within the framework of supersymmetric
quantum mechanics is necessary before proceeding.

\section{Application to the $^{11}Be$ system}

Research with radioactive nuclear beams is currently one of the most active
areas in nuclear physics. As one of the successful applications of such
nuclear beams, exotic structures have been observed in nuclei near to the
driplines, which are called, as the most interesting discoveries, the
neutron halo. These nuclei have opened studies of weakly bound nuclear
systems, which has not freely accessible before. These nuclei, such as $%
^{11}Be$, have long-range wave functions and are characterized by a cloud,
or halo, of neutron probability that extends far beyond the dense core.
According to classical physics such nuclei should not exist at all because
the strong nuclear force (the glue that binds neutrons and protons together)
has too short to hold the far off neutrons in the halo. Instead, they owe
their existence to quantum theory which describes the location of subatomic
particles by a mathematical cloud of probability. In \cite{Zhukov} and
references therein essential features of loosely bound systems, having an
unusually large size, are discussed. In this Section, as an example to halo
systems, we examine the ground state of $^{11}Be$, which consists of a
single neutron halo with a $^{10}Be$ core nucleus, using the supersymmetric
quantum mechanics. It is well known that the dominant component of the $%
^{11}Be$ ground state is produced by the coupling of a $1s_{1/2}$ neutron,
having the separation energy of $0.503$ $MeV$, to a $^{10}Be$ core.

\subsection{Phase-equivalent potentials for the $^{11}Be$ system}

In this analysis, which will be restricted to the {\it s}-motion only, we
again make use of the $\sec $h-squared deep potential with appropriate
physical parameters. At this stage we should stress that one can relate the
matter root-mean-square radius $R_{RMS}(matter)$ to the single neutron
root-mean-square radius $R_{rms}$ by the formula \cite{Zhukov}
\begin{equation}
R_{RMS}^{2}(matter)=\frac W{\left( W+1\right)
}R_{RMS}^{2}(core)+\frac W{\left( W+1\right) ^2}R_{rms}^{2}
\end{equation}
where $W$ is the mass number of the core, here the $^{10}Be$
nucleus. Using \cite{Van Dijk} the $R_{RMS}(core)=2.3$ $fm$ and
$R_{RMS}(matter)=2.73$ $fm$, it is easy to check that the value
$R_{rms}=6.70$ $fm$ gives roughly the average of the measured
values of $R_{RMS}(matter)$.

To determine the two free parameters (${\tilde{A}}$, $\beta $ ),
one needs to solve the following equations leading to the correct
$rms$ value and binding energy for the $Be-11$ system,

\begin{eqnarray}
E^{(n)} &=&-\frac{\hbar ^{2}}{2\mu }(\tilde{A}-2n-1)=-0.503~MeV;
\\
R_{rms}^{2} &=&\int drr^{2}\Psi _{m}^{(n)}(\tilde{A},\beta
,r)^{2}=(6.70~fm) \nonumber
\end{eqnarray}
where $\mu$ is the reduced mass of the system: $^{10}Be+n$.

The physical solution is chosen using the node-number and parity quantum
number criteria. As we take into account that $0s_{1/2}$ the orbit is
completely occupied and put the single neutron in $1s_{1/2}$ state (having
one node in the wave function) to have a positive parity required by the
experiment, we choose the excited state wave function with one node, instead
of ground state, as a physically meaningful solution for the superdeep
potential in analyzing the ground state of $^{11}Be$ nucleus.

Now we are ready to apply the supersymmetric technique to create the PEP
corresponding to the $\sec $h-squared superdeep potential for the $^{11}Be$
one-nucleon halo system, with the calculated values of ${\tilde{A}%
=3.124}$, and $\beta =0.694$ $fm^{-1}$ together with $\frac{\hbar ^{2}}{2\mu }%
=22.81$ $MeV$ $fm^{2}$. Using the formulae given by the previous
sections, the supersymmetric partner potentials and corresponding
eigenstates are calculated. Fig. 3 illustrates the superdeep and
related partner potentials while Fig. 4 gives their eigenstates
respectively. It is seen from Fig. 4 that the eigenfunction
corresponding to the PEP is a nodeless ground state. In this case
the Pauli principle is taken into account by the repulsive part of
the potential, see Fig. 3, repulsive up to $1.5$ $fm$
approximately.

If one calculates the $R_{rms}$ for the system of interest using the
nodeless eigenstate of PEP, the value of $6.78$ $fm$ is obtained. Even the
wave functions differ in the node number, like the deuteron case discussed
in Section 3, the $R_{rms}$ is nearly the same for the phase-equivalent
potentials, which is found about $6.70$ $fm$ for the initial deep potential,
whereas $6.17$ $fm$ for the intermediate non-phase-equivalent potential. One
should not forget that the non-PEP potential does not belong to the PEP
family, that is why the corresponding $R_{rms}$ values so close for the PEP
potentials while that of non-PEP considerably differ than the others.
Moreover, the corresponding eigenfunction of this non-PEP potential has a
different asymptotic behavior as well.

We have also calculated the {\it s}-wave phase shifts in case of $%
^{10}Be(d,p)^{11}Be$ elastic scattering up to $20$ $MeV$ for the three
potentials. The results clearly exhibit the difference between PEP and
non-PEP.

\section{Application to the $^{11}Be(p,d)^{10}Be$ reaction at 35 MeV}

Single nucleon transfer reactions, such as the $(d,p)$ and $(p,d)$
reactions, have been a reliable tool in nuclear spectroscopic studies of
stable nuclei, determining positions, spins and parities of nuclear states.
Recently, the use of low energy single nucleon transfer reactions for
structure studies of exotic nuclei have attracted attention \cite{Winfield}-%
\cite{Lenske}. Because of the simplicity of the theoretical interpretation
of these reactions, they are thought to provide an important source of the
information about the structure of halo nuclei, such as $^{11}Be$. It is now
understood that the $1s_{1/2}$ neutron single particle state in this region
is lowered and that a dominant component of the $^{11}Be$ ground state is
produced by the coupling of a $1s_{1/2}$ neutron to a $^{10}Be$ (g.s., $%
0^{+} $) core; with a smaller but significant component in which a
$0d_{5/2}$ neutron is coupled to a $2^{+}$ excitation of the
$^{10}Be$ core.

The importance upon such transfer reaction spectroscopic studies of the
inclusion of the deuteron breakup degrees of freedom has been well discussed
in Ref. \cite{Lenske} via the theories used to analyze measured cross
section observables, and shown that the magnitudes of the calculated cross
sections, and particularly the ratio of the cross sections to the ground
state and $2^{+}$ core states, of $^{10}Be$ are affected by the inclusion of
three-body channels.

Here in this section we do not discuss the details of these calculations.
The aim of the present calculations is to investigate how the calculated
physical observables of the reaction sensitive to the nodal structure,
involving the deuteron and bound neutron wave functions discussed in the
previous sections with one node/no node.

We calculate the transfer amplitude using the prior form of the $(p,d)$
matrix element, thus the transition interaction is the $n-p$ interaction and
we need a full (three-body) description of the $n+p+$ $^{10}Be$ system in
the final state. For the description of this final state we have used both
the adiabatic (AD) model \cite{Johnson} and the quasi-adiabatic (QAD)
approach \cite{Amakawa}. However, for the sake of clarity in discussing the
physics behind this application, we here consider only the AD model
calculations as both model calculation results have led us to the same
conclusion.

In the context of $(p,d)$ reactions the outgoing deuteron wave function, $%
\Psi _{d}$, enters the transition amplitude

\begin{equation}
T_{pd}=\left\langle \Psi _{d}(\overline{r},\overline{R})\right|
V_{np}\left| \chi _{p}(\overline{r}_{p})\phi
_{n}(\overline{r}_{n})\right\rangle~,
\end{equation}
where $\overline{r}$ ($=\overline{r}_{p}-\overline{r}_{n}$) is the
relative coordinate of the $n-p$ pair and $\overline{R}\left[
=\frac {1}{2}\left( \overline{r}_{p}+\overline{r}_{n}\right)
\right] $ is the center of mass coordinate. Here $\phi _n$ is the
neutron bound state and $\chi _{p}$ incoming proton wave function.
The vectors $\overline{r}_{n}$ and $\overline{r}_{p}$ are the
positions of the transferred neutron with respect to the $^{10}Be$
core and of the proton relative to the $^{11}Be$ system. The cross
section (in the center of mass frame) for neutron pickup to bound
$n-p$ pair is then given by
\begin{equation}
\frac{d\sigma _{pd}}{d\Omega _{f}}=\frac{\mu _{i}\mu _{f}}{\left(
2\pi \hbar ^{2}\right) ^{2}}\frac{k_{f}}{k_{i}}\left|
T_{pd}\right| ^{2}
\end{equation}
where $d\Omega _{f}$ is the element of the solid angle for the
asymptotic
center of mass momentum of the bound deuteron. In the above $\mu _{i}$ and $%
\mu _{f}$ are the center of mass relative motion reduced masses in
the initial and final channels, $k_{i}$ and $k_{f}$ are the
entrance and exit channel wave
numbers, respectively. Throughout this paper we restrict the formalism to $s$%
-wave $n-p$ relative motion for simplicity. In zero-range approximation then
it is the wave function at coincidence, $\Psi _{d}$ $\left( r\approx 0,%
\overline{R}\right) $ which is of importance.

\subsection{Transition amplitudes in zero-range approximation}

It is important for the present work to make clear the essential
differences in the calculations carried out using both
phase-equivalent potentials, the initial superdeep $sech$-squared
potential and the shallow phase-equivalent partner potential, in
calculating the ground state deuteron wave function in the final
state and for the calculation of the transferred neutron bound
state wave function in the initial channel of the reaction. So
they are developed in some detail within a common notation. The
calculations employed the deep potential (denoted by $V_{1}$ in
sections through $2-4$) will be represented by the script $deep$
in the following formulae and for the phase-equivalent shallow
potential (represented by $V_{3}$ earlier) calculations we use the
script $pep$.

The transition amplitudes for the processes are evaluated in a zero-range
approximation, and the related relative amplitudes, of primary interest
here, are accurately described. For clarity we will not show the transferred
neutron spectroscopic factor or any spin projection labels explicitly.

In the adiabatic approximation the required transition amplitudes are
\begin{equation}
T_{pd}^{deep(pep)}=\left\langle \chi _{d}(r,\overline{R})\Phi
_{d}^{deep(pep)}(r)\right| V_{np}^{deep(pep)}(r)\left| \chi _{p}(\overline{r}%
_{p})\phi _{n}^{deep(pep)}(\overline{r}_{n})\right\rangle
\end{equation}

As in calculating the transfer amplitudes, we make use of the zero-range
approximation, thus for the bound deuteron we replace
\begin{equation}
V_{np}^{deep(pep)}(r)\chi _{d}(r,\overline{R})\phi
_{d}^{deep(pep)}(r)\approx D_{0}^{deep(pep)}(p,d)\chi
_{d}(r\approx 0,\overline{R})\delta (r)
\end{equation}
with strength parameters
\begin{equation}
D_{0}^{deep(pep)}(p,d)=\sqrt{4\pi }\int
drrV_{np}^{deep(pep)}(r)u_{0}^{deep(pep)}(r)~,
\end{equation}
where $u_{0}$ is the radial deuteron ground state wave function ($\ell _{np}=0$%
) such that $\Phi _{d}(r)=\frac {1}{\sqrt{4\pi
}}\frac{u_{0}(r)}{r}$. The similar replacement should also be done
for the QAD calculations.

It follows that,
\begin{eqnarray}
T_{pd}^{deep(pep)} &=&D_{0}^{deep(pep)}(p,d)M^{deep(pep)}(p,d) \\
&=&D_{0}^{deep(pep)}(p,d)\left\langle \chi _{d}(r\approx 0,\overline{R}%
)\right| \left. \chi _{p}(\gamma \overline{R})\phi _{n}^{deep(pep)}(\overline{%
R})\right\rangle   \nonumber
\end{eqnarray}
where $\gamma =\frac {W}{W+1}$. Finally, considering the general
expression for the cross section, we obtain
\begin{equation}
\frac{d\sigma _{pd}^{deep(pep)}}{d\Omega _{f}}=\frac{\mu _{i}\mu
_{f}}{\left( 2\pi \hbar ^{2}\right)
^{2}}\frac{k_{f}}{k_{i}}D_{0}^{2}(deep,pep)\left|
M^{deep(pep)}(p,d)\right| ^{2}
\end{equation}
for the calculations used the adiabatic model in describing the final state.

\subsection{Calculation methods}

We calculate the cross section angular distributions for the $%
^{11}Be(p,d)^{10}Be$ single nucleon transfer reaction leading to the $0^{+}$
ground state ($1s_{1/2}$ neutron transfer) of $^{10}Be$. We perform
zero-range calculations using a modified version of the program TWOFNR \cite
{Igarashi}. The program has been further modified so that the calculated
adiabatic and quasi-adiabatic radial wave functions, and appropriate
zero-range strengths calculated for the phase equivalent deep and shallow
potentials by Eq. (41), can be read in, together with the transferred
neutron bound state $\phi _{n}$ obtained using the superdeep $sech$%
-squared potential and its phase-equivalent partner. In the
three-body model calculations of the deuteron channel wave
function ($d+^{10}Be$), we make use of the global optical
potential parameter set of Bechetti and Greenlees \cite{Bechetti}
evaluated at half the final state center of mass energy. The
spin-orbit interactions are included. The entrance channel proton
optical potential parameters are taken from \cite{Bechetti}. The
spin-orbit force in the proton channel is fixed at $6$ $MeV$. The
radial integrals are carried out from $0$ to $35$ $fm$ in steps of
$0.1$ $fm$. The maximum number of partial waves used was $30$ for
both the entrance and exit channels. The spectroscopic factors are
set to unity throughout the calculations. All calculations
presented here are done without non-locality corrections. Such
corrections for halo transfer are expected to be small because
they correct the transition amplitude in the nuclear interior, but
the long tail of the halo wave function makes internal
contributions less important.

\subsection{Results and discussion}

It is useful at this stage to remind ourselves again that the deep potential
and its phase equivalent shallow partner, which are used for calculating the
bound deuteron and transferred neutron wave functions, are constructed to
have identical phase shifts so that any difference in the transition
amplitudes, and in the cross section angular distributions, is attributed
entirely to the corresponding wavefunctions of the partner potentials.

The result for the adiabatic model cross section angular distributions for
the reaction involving the original deep potential description and its
comparison with that obtained by means of the PEP description is given in
Fig. 5. The figure indicates almost complete coincidence of both curves.
This may be understood from the following analysis. The two-body
supersymmetric partner potential dependence in the differential cross
section calculations, originates from two terms: the zero-range constant $%
D_0^2$, and the transition amplitude $\left| M(p,d)\right| ^{2}$
through the deuteron and bound neutron ground state wave
functions. The calculated transfer strengths, for the deep
potential description of the $n-p$ interaction is
$D_{0}^{2}(deep)=15792$ $MeV^{2}$ $fm^{3}$ while
$D_{0}^{2}(pep)=15980$ $MeV^{2}$ $fm^{3}$ for the shallow partner
description. It is also worth stressing that, from the results
obtained in sections 3 and 4, the transition from the deep
potential to the repulsive core interaction (PEP) does not
significantly alter the outer part of the neutron ground state
wave functions appear in the transition amplitude. And in
addition, the neutron
halo wave function makes internal contribution less important. Therefore $%
\left| M^{deep}(p,d)\right| ^{2}\approx \left| M^{pep}(p,d)\right|
^{2}$, and the ratio of the cross sections
\begin{eqnarray}
\frac{d\sigma _{pd}^{deep}}{d\Omega _{pd}^{pep}} &\approx &\frac{D_{0}^{2}(deep)%
}{D_{0}^{2}(pep)}=\frac{15792~MeV^{2}fm^{3}}{15980~MeV^{2}fm^{3}}
\\ &=&0.988\approx 1~, \nonumber
\end{eqnarray}
which leads to the coincidence of the results.

\section{Conclusion}

The properties of the deep nuclear interaction have been investigated by
constructing explicitly phase-equivalent potentials freed from the
unphysical deeply bound states of the former. We have seen that the
resulting central potentials have to be repulsive and singular at small
distance in order to preserve the energy behavior of the phase shifts, and
they present a shallow attractive part of intermediate range. Our
reconstructed potentials (PEP) have led to relative motion wave functions
very similar to those generated by the deep potentials outside the core
region, but which lack the small distance radial node. Both types of
potentials are therefore expected to display rather different off-shell
behaviors, and presumably lead to qualitatively different results. However,
there is no considerable discrepancy between the $rms$ calculation results
of these quite different two-body interaction descriptions has been found.
Nevertheless, if experiment actually does require a reduction in observables
then a short range non-local contribution to the potential, like the deep
potential, appears to be necessary.

We have used phase equivalent two-body potentials with a different
number of bound states considering the $^{11}Be(p,d)^{10}Be$
reaction at $35$ $MeV$, and compared the calculated corresponding
physical observables. Investigation of the consequences of using
these completely phase equivalent two-body potentials for the
description of weakly bound deuteron and $^{11}Be $ nuclei in
three-body calculations, based on the adiabatic approach has led
us to almost indistinguishable results. Due to the large spatial
extension of halos, involving the simplest halo nucleus the
deuteron and $^{11}Be$ as a one-neutron halo system, the
probability is by definition very small at small distances. Hence,
we conclude that the short range behavior of the corresponding
wavefunctions for the deep and phase-equivalent shallow
potentials, which coincide at large distances but differ at small
distances by the additional node appearing inside the core by use
of the deep potential, is not significant for the analysis of such
reactions.

In sum, the supersymmetric formulations used through the present
calculations have dealt in general with the Pauli principle for the weakly
bound systems. Assuming that the two-body potentials have Pauli forbidden
states, one can then construct easily and use the phase equivalent partners
without these forbidden states. At small distances the lowest levels of the
original deep potentials correspond to identical fermions occupying the same
states. Removal of these terms therefore forces the particles to occupy
higher-lying orbits and thereby introducing the necessary repulsion
preventing violation of the Pauli principle. In conclusion, this method to
exclude the Pauli forbidden states in the weakly bound systems has firm
mathematical and numerical foundations. It is a practical and accurate
alternative to the other existing methods, such as the work described in
Ref. \cite{Garrido} where an analytical $s$-wave potential with one bound
state, which is the most important case in the practical applications for
halo states, has been introduced. We note that the application of this
potential, with the appropriate choice of the parameters involved, to the
weakly bound deuteron and $^{11}Be$ nuclei has led to the similar results to
those obtained by the shallow supersymmetric phase-equivalent partner
potentials.

The valuable discussions with D. Baye, W. Van Dijk, and M. W. Kermode are
here gratefully acknowledged.

\newpage\

\newpage\

\begin{center}
{\bf Figure Captions }
\end{center}

{\bf Fig. 1-a.} Superdeep potential $V_{1}(r)$ (solid line) for
the deuteron
system and its supersymmetric partners $V_{2}(r)$ (non-PEP, dotted line) and $%
V_{3}(r)$ (PEP, dashed line) as a function of radius $r$.

{\bf Fig. 1-b.} Comparison of the $n-p$ central potential of the
superdeep $sech$-squared $V_{1}(r)$ (solid line) potential and its
PEP potential $V_{3}(r)$ (dashed line) with the central Reid Soft
Core (dotted line) interaction.

{\bf Fig. 2.} The first two eigenstates, $n=0$ for the ground
state (solid line) and $n=1$ for the first excited state (dotted
line) of the original hamiltonian with the superdeep two-parameter
potential for the deuteron. The wave function illustrated by
dashed line represents the ground state of the SUSY PEP,
$V_{2}(r)$.

{\bf Fig. 3.} The same as Fig.1-a, but calculations have been carried out
for the $^{11}Be$ system.

{\bf Fig. 4.} The $^{11}Be$ ground-state radial wave functions $\Psi
_{1}^{(n=1)}({\tilde{A}},\beta ,r)$ (solid curve), $\Psi _{2}^{(n=0)}(%
{\tilde{A}},\beta ,r)$ (dotted curve) and $\Psi _{3}^{(n=0)}(%
{\tilde{A}},\beta ,r)$ (dashed curve) all normalized to one.

{\bf Fig. 5.} Calculated differential cross section angular distributions
within the adiabatic model for the $^{11}Be(p,d)^{10}Be$ (g.s.) reaction at $%
35$ $MeV$ using the deep and shallow (PEP) two-body potential descriptions
for the weakly bound deuteron and $^{11}Be$ nuclei.

\end{document}